\documentclass[aps,preprint,showpacs]{revtex4}
\usepackage{graphicx,epsfig}
\usepackage{amssymb}
\begin{document}

\title{Frequency Shifts of photons emitted from geodesics of nonlinear electromagnetic black holes.}

\author{L. A. L\'opez $^{1}$ }
\email{lalopez@uaeh.edu.mx}
\author{J. C. Olvera $^2$}
\email{jcolvera@fis.cinvestav.mx}

\affiliation{$^1$  \'Area Acad\'emica de Matem\'aticas y F\'isica.,  UAEH, carretera Pachuca-Tulancingo km 4.5, C.P. 42184 , Pachuca, Hidalgo, M\'exico}
\affiliation{$^2$ Dpto de F\'isica, Centro de Investigaci\'on y de Estudios Avanzados del I.P.N,
14-740 Apdo, DF, M\'exico}

\begin{abstract}
We analyzed the frequency shifts of photons emitted from massive particles that move in stable geodesics around black holes with non-linear electrodynamics. The motion of the photon is modified by an effective metric when the  Einstein gravity is coupled with non-linear electrodynamics. The kinematic shifts for photons are modified by the magnetic and electric factors that make the difference between the linear and non-linear electrodynamics.
As an illustration, we present the frequency shifts of the photons for black holes with non-linear electrodynamics, two magnetic black holes, and two electric black holes,  the Bardeen and Bronnikov black holes for the magnetic charge and the Born-Infeld and Dymnikova black holes for electric charge. These frequency shifts are compared with their linear electromagnetic counterpart.
\end{abstract}

\pacs{04.70.Bw,04.70.-s, 42.15.-i, 05.45.-a}

\maketitle

\section{Introduction}

The evidence that several galaxies contain a supermassive black hole at their center, with a mass ranging from millions to billions of solar masses \cite{Begelman1898}, has led to various studies and observations. For example, different studies have provided strong evidence that Sgr A* (an extremely compact radio source at the center of our Galaxy) is a supermassive black hole \cite{PMID:16267548}.

Shortly project GRAVITY \cite{2009ASSP....9..361E},  will track the stars orbiting Sgr A*, then it needs to propose models or novel relativistic methods for determining parameters corresponding to black holes in terms of directly observed magnitudes.

In \cite{Herrera-Aguilar:2015kea} a theoretical approach was developed to obtain the parameter of a Kerr black hole in terms of the redshift and blueshift of photons emitted by massive particles traveling along stable circular geodesics around a Kerr black hole. Using this idea, different configurations of black holes have been studied. In  \cite{Becerril:2016qxf} the  Reissner-Nordstr\"{o}m (RN) black hole was studied. Also in \cite{Kraniotis:2019ked}, the red-blueshift for Kerr-Newman-de Sitter and Kerr-Newman black holes are studied.

It is well known that in the presence of non-linear electrodynamics (NED), the behavior of photons and massless test particles are not the same, the photons propagate along null geodesics of an effective geometry that depends on the non-linear theory. Then, the electromagnetic fields modify the trajectories of the null geodesics \cite{Gutierrez:1981ed}. The effects of no-linear electrodynamics have been analyzed by various authors to study a variety of physical situations, for example, the quasinormal modes of non-linear electromagnetic black holes \cite{Breton:2016mqh}, the Scattering and absorption sections black hole in NED \cite{Olvera:2019unw} and the shadow of Bardeen black hole in NED \cite{Stuchlik:2019uvf} as well as the shadow of  Euler-Heisenberg  and  Bronnikov black holes \cite{Allahyari:2019jqz}.

Also in \cite{Cuzinatto:2015xta}, the non-linear electrodynamics effects on the geodesic deviation and the redshift of photons propagating near spherically symmetric mass and charge distributions are analyzed  up to the first-order considering the effective metric.

The paper is organized as follows: Sec. II shows how to determine frequency shifts of photons emitted from massive particles that move in stable geodesics around black holes with NED, considering the modifications of the photon trajectories. At the end of this section we give the modified expression of the redshifts in term of the magnetic and electric factors of the effective metric.
In Section III we analyze the frequency shifts of the four examples mentioned above. In each case, the frequency shifts are compared with the ones corresponding to the massless test particles and light rays of the Reissner-Nordstr\"{o}m (RN) black hole that is the linear counterpart of NED. Finally, conclusions are given in the last section.

\section{The red-blueshifts for photon}

The connection between the red-blueshifts of the photons emitted by massive particles that move around of a black hole was pointed out in\cite{Herrera-Aguilar:2015kea}. In this section we give a summary of the method considering a statistical space-time as in \cite{Becerril:2016qxf} and we introduce the modifications caused by the NED.

First, we consider the spherical coordinates $x^{\mu}=(t,r,\theta,\phi)$ and for a static  spherically symmetric background;

\begin{equation}\label{sss}
ds^{2}=g_{\mu \nu}dx^{\mu}dx^{\nu}=-f(r)dt^{2}+\frac{1}{g(r)}dr^{2}+r^{2}d\Omega^{2},
\end{equation},  

 the particles propagate in the space-time following the geodesic equation;

\begin{equation}\label{geo}
\delta=g_{\mu \nu}\dot{x}^{\mu}\dot{x}^{\mu}=-f(r)\dot{t}^{2}+\frac{1}{g(r)}\dot{r}^{2}+r^{2}\dot{\theta}^{2}+r^{2} \sin \theta \dot{\phi}^{2}
\end{equation}

where $\dot{x}^{\mu}=U^{\mu}=\frac{dx^{\mu}}{d\tau}$ is the $4-$velocity, $\tau$ is an affine parameter. When  $\delta=0$, the equation corresponds to photon, while $\delta=-1$ represents massive particles.

The conserved quantities of the massive particles are the energy $E$ and angular momentum $L$, given by;

\begin{equation}\label{U}
U^{t}=-\frac{E}{g_{tt}}=\frac{E}{f(r)} \;\;\;\;\;\;\;\;\;\ U^{\phi}=\frac{L}{g_{\phi\phi}}=\frac{L}{r^{2}\sin\theta}
\end{equation}

For equatorial orbits $(\theta=\pi/2)$, the equation of motion for a test particle in the static space-time is given by $g_{rr}\dot{r}^{2}+V_{eff}=0$, where $V_{eff}$ is the effective potential for radial motion;

\begin{equation}\label{EfectiveV}
V_{eff}=1+\frac{E^{2}g_{\phi\phi}+L^{2}g_{tt}}{g_{\phi\phi}g_{tt}}=1-\frac{E^{2}r^{2}-L^{2}f(r)}{r^{2}f(r)}.
\end{equation}

For  the geodesics of the massive particles (stars or gas) to be  considered as circular orbits stable they must fulfill the following conditions; $V_{eff}^{'}(r_{c}) = V_{eff}(r_{c})=0$ where $r_{c}$ is the radius of the circular orbit and the prime denotes derivative respect to $r$. From these two conditions it is possible to write the constants of motion as;

\begin{equation}\label{E}
E^{2}=-\frac{g_{tt}^{2}g_{\phi\phi}^{'}}{g_{tt}g_{\phi\phi}^{'}-g_{tt}^{'}g_{\phi\phi}} =\frac{2f^{2}(r)}{2f(r)-rf^{'}(r)}\left.\right|_{r_{c}}
\end{equation}

\begin{equation}\label{L}
L^{2}=\frac{g_{\phi\phi}^{2}g_{tt}^{'}}{g_{tt}g_{\phi\phi}^{'}-g_{tt}^{'}g_{\phi\phi}}=\frac{r^{3}f^{'}(r)}{2f(r)-rf^{'}(r)}\mid_{r_{c}}
\end{equation}

Finally for circular stable orbits to exist, the condition $V_{eff}^{''}(r_{c})>0$ must be fulfilled.

The aim is to get the red $z_{r}$ and blue shifts $z_{b}$ of the light emitted by stars or gas orbiting around of the black hole. The emitted photons have $4-$momentum $\kappa^{\mu}$ and move along null geodesics $\gamma_{\mu \nu}\kappa^{\nu}\kappa^{\mu}=0$, where $\gamma_{\mu \nu}$ is the effective metric.

The calculation of the effective metric components $\gamma_{\mu\nu}$ of an static space-time is shown in \cite{Breton:2016mqh}. The line element is given by;

\begin{equation}\label{efec}
dS_{\rm eff}^{2}=\gamma_{\mu \nu}dx^{\mu}dx^{\nu}=(\mathcal{L}_{F}G_{m})^{-1}\left\{G_{m}G_{e}^{-1}\left( -f(r)dt^{2} + \frac{1}{g(r)}dr^{2}\right)+r^{2}d\Omega^{2}\right\}.
\end{equation}

where $G_m$ and $G_e$ are the magnetic and electric factors that make the difference between the linear and non-linear electromagnetism and are given by;

\begin{equation}\label{Gs}
G_{m}=\left(1+4\mathcal{L}_{FF}\frac{q_{m}^{2}}{\mathcal{L}_{F}r^{4}}\right) , \;\;\;\;\;\;\ G_{e}=\left(1-4\mathcal{L}_{FF}\frac{q_{e}^{2}}{\mathcal{L}_{F}^{3}r^{4}}\right);
\end{equation}

$\mathcal{L}$ is an arbitrary function of the electromagnetic invariant  $F=F^{\mu\nu}F_{\mu\nu}$ and the subindex $F$ represents the first or second derivative with respect to $F$. $G_{m}$ and $G_{e}$ in the linear limit become equal to one.

 The action for  gravitation  coupled to  a non-linear electrodynamics (NED) is;

\begin{equation}\label{action}
S=\frac{1}{16\pi}\int d^{4}x\sqrt{-g}[R-\mathcal{L}(F)],
\end{equation}

where $R$ is the scalar curvature. 

The energy $-E_{\gamma}=\gamma_{tt}\kappa^{t}$ and angular momentum $L_{\gamma}=\gamma_{\phi\phi}\kappa^{\phi}$ of the emitted photon are conserved quantities. 

The frequency shift $z$ associated with the emission (e) and detection (d) of photons is defined as;

\begin{equation}\label{Z1}
1+z=\frac{\omega_{e}}{\omega_{d}}=\frac{-\kappa_{\mu}U^{\mu}\mid_{e}}{-\kappa_{\nu}U^{\nu}\mid_{d}}
\end{equation}

where $\omega_{e}$ is the frequencies emitted by an observer moving in the circular stable orbit and $\omega_{d}$ the frequency detected. 

If the detector (or observer) is located very far away from the source $(r \rightarrow \infty)$, $U^{t}=E=1$. In the case of circular orbits, $U^{r}=0$  and in the equatorial motion $U^{\theta}=0$, equation (\ref{Z1}) takes the form;

 \begin{equation}
1+z=\frac{(E_{\gamma}U^{t}-L_{\gamma}U^{\phi})\mid_{e}}{(E_{\gamma}U^{t}-L_{\gamma}U^{\phi})\mid_{d}}=\frac{U^{t}_{e}-b_{e}U^{\phi}_{e}}{U^{t}_{d}-b_{d}U^{\phi}_{d}},
\end{equation}

where $b$ is  the impact parameter of the photons $b=L_{\gamma}/E_{\gamma}$ and with $b_{e}=b_{d}$ . Now it is necessary to consider the kinematic redshift of photons  $z_{kin}$  along the line of sight that links the black hole and the observer. Then for a $1+z_{c}=U^{t}_{e}/U^{t}_{d}$ redshift evaluated at the central value ($b=0$)  the kinematic redshift of photons is defined as $z_{kin}=z-z_{c}$.
 
From $\kappa_{\nu}\kappa^{\nu}=0$ with $\kappa^{r}=\kappa^{\theta}=0$, it is possible to obtain $b=\pm \sqrt{-\frac{\gamma_{\phi\phi}}{\gamma_{tt}}}$. There are two different values of $b$  ($b_{+}=-b_{-}$),  these values represent the photons emitted by a receding $z_{1}$ (redshift) or an approaching object $z_{2}$ (blueshift) with respect to a distant observer i.e the kinematic shifts of photons emitted from either side of the central value read as $|z_{1}|=| -z_{2} |$.

If the detector is located far away from the object, $r_{d} \rightarrow \infty$ and $U_{d}^{\mu} \rightarrow (1,0,0,0) $ result in;

\begin{equation}\label{z1}
z_{1}=U_{e}^{\phi}b_{e_{+}}.
\end{equation}

Now, using the equations (\ref{U}), (\ref{E})and (\ref{L}), the expression for the $4-$ velocity $U^{\phi}_{e}$ in terms of the metric components is;

\begin{equation}
U^{\phi}_{e}=\sqrt{\frac{g_{tt}^{'}}{g_{tt}g_{\phi\phi}^{'}-g_{tt}^{'}g_{\phi\phi}}}=\sqrt{\frac{f^{'}(r)}{2rf(r)+r^{2}f^{'}(r)}}.
\end{equation}

Then the explicit form of (\ref{Z1}), is given by:

\begin{equation}\label{redblue}
z_{ph}=z_{1}=\sqrt{\frac{1}{G_{m}G_{e}^{-1}}\frac{rf^{'}(r)}{f(r)\left(2f(r)+rf^{'}(r)\right)}}=\sqrt{\frac{G_{e}}{G_{m}}}z_{massless}
\end{equation}

Where $z_{massless}$ is linear counterpart ($G_{e}=G_{m}=1$) of $z_{ph}$.  $z_{ph}$ is enhanced or suppressed as compared to the linear counterpart $z_{massless}$ depending of the arbitrary function $\mathcal{L}(F)$.
For example, if $\mathcal{L}_{FF}/\mathcal{L}_{F} < 0$ then $G_{m} \leqslant 1$ and $G_{e}> 1$ and consequently  $z_{ph}$ gets a mayor value compared to the linear counterpart.

\section{Examples}

As an illustration we calculate the kinematic shifts for four black holes in NLED theory, considering the modification of the effective metric and the impact parameter. Then, we analyze and compare with the case when the effective metric is not considered (for massless particles). Also the comparison with the kinematic shifts of the linear electromagnetic counterpart (RN black hole) is done.

\subsection{Bardeen black hole}

The Bardeen BH can be interpreted as a self-gravitating  nonlinear magnetic monopole \cite{AyonBeato:2000zs} with mass $M$ and magnetic charge $q_{m}=g$,  derived from  Einstein gravity coupled to the nonlinear Lagrangian,

\begin{equation}
\mathcal{L}(F)=\frac{6}{s g^{2}}\frac{(g^{2}F/2)^{\frac{5}{4}}}{(1+\sqrt{g^{2}F/2})^{\frac{5}{2}}},
\label{ABGlag}
\end{equation}

where  $s= |g|/2M$.  The solution for the coupled Einstein  and NLED Lagrangian (\ref{ABGlag}) for a static spherically symmetric space  is given by,

\begin{equation}\label{Bardeenmetric1}
f(r)= g(r)=1-\frac{2Mr^{2}}{(r^{2}+g^{2})^{\frac{3}{2}}}.
\end{equation}

The solution (\ref{Bardeenmetric1}) has horizons only if $2s= g/M \leq \frac{4}{3\sqrt{3}}$. The electromagnetic invariant is $F=2g^{2}/r^{4}$ and the $G_e$ and $G_m$ factors are  given by

\begin{equation}
G_{e}=1, \quad G_{m}=1-\frac{4(6g^{2}-r^{2})}{8(r^{2}+g^{2})}.
\label{GsBardeen}
\end{equation}

\begin{figure}[h]
\begin{center}
\includegraphics [width =0.45 \textwidth ]{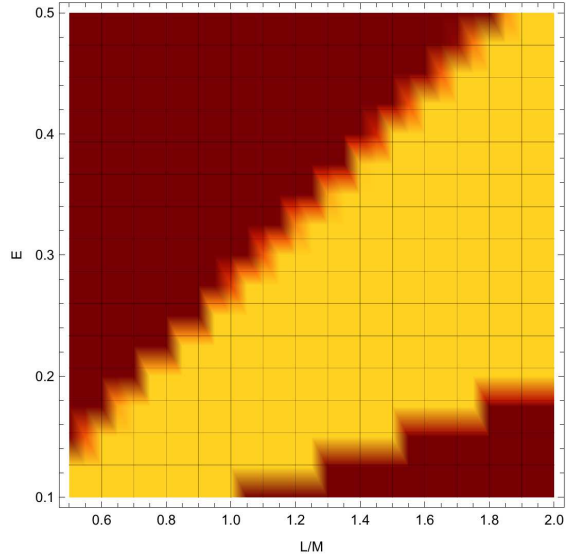}
\end{center}
\caption{Density plot for the parameters $E$ and $L/M$ in Bardeen BH. Lighter regions represent a higher number of possible orbits.}\label{Fig1}
\end{figure}

The circular orbits on the equatorial plane exist provided that the equation (\ref{E}) ($E^{2}>0$) and the equation (\ref{L}) ($L^{2}>0$) are simultaneously fulfilled. Also the stability of these circular orbits requires that the $V_{ef}^{''}(r_{c})>0$. This analysis is performed numerically, figure (\ref{Fig1}), shows a density plot which illustrates the combinations for $E$ and $L/M$ that allow for a greater number of circular orbits (represented by the lighter part of the plot).

\begin{figure}[h]
\begin{center}
\includegraphics [width =0.5 \textwidth ]{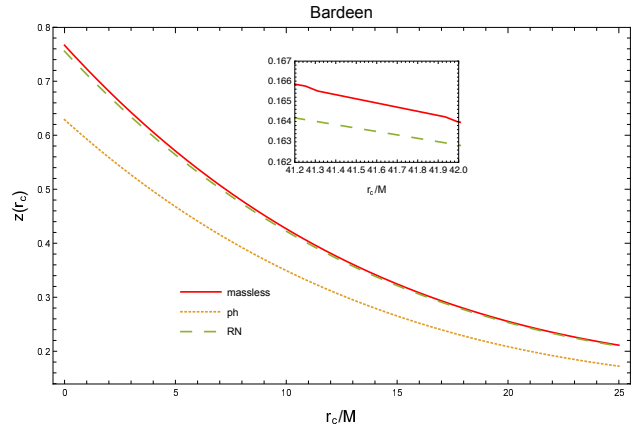}
\end{center}
\caption{Behavior of kinematic redshift for  Bardeen black hole with values $E=0.5$ and $L/M=2$.}\label{Fig2}
\end{figure}

It is possible to obtain the behavior of $z$, which is shown in figure (\ref{Fig2}) for different values of $r_{c}/M$. The value of $z$ for Bardeen black hole, decreases as $\frac{r_{c}}{M}$ augments. 

It can be noted that when massless particles are considered, the behavior of $z$ redshift in function of $r_{c}/M$ for Bardeen black hole remains near RN black hole as can be noted from  (\ref{Fig2}), while the deviation with respect to the photon (ph) case is more significant ($ z_{ph}< z_{RN}<z_{massless}$).
 
\begin{figure}[h]
\begin{center}
\includegraphics [width =0.5 \textwidth ]{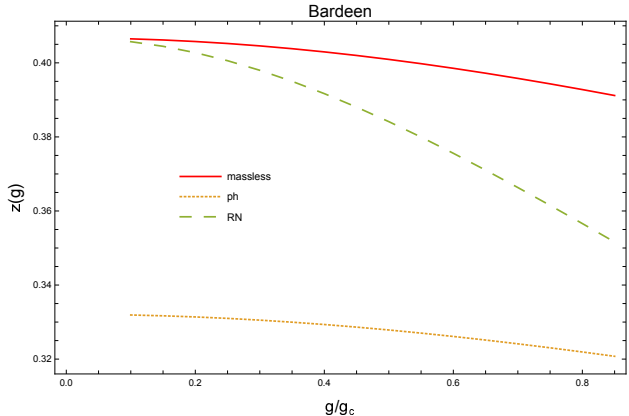}
\end{center}
\caption{Behavior of kinematic redshift for  Bardeen black hole with values $E=0.5$ and $L/M=2$.}\label{Fig3}
\end{figure}

To carry out the study, we define the critical charge $g_{c}= \frac{4M}{3\sqrt{3}}$. Then the $z$ redshift in terms of $g/g_{c}$ for Bardeen black hole, from massless particles and photons (ph) are compared in Fig (\ref{Fig3}), where also the redshift for RN is shown. Then it is possible to observe that  $ z_{ph}< z_{RN}<z_{massless}$.

The resemblance with the $z$ of RN occurs only when the charge approaches zero in the case of massless particles. However, in the case of photons in the limit that charge goes to zero, the RN behavior is not recovered. Note that for $g \mapsto 0$ the correct limit $G_m=1$ is not achieved, instead $G_m(g \mapsto 0)= 3/2$,  \cite{Breton:2016mqh} gives one possible explanation, that may be the nature of the solution because the charge and mass parameters are not independent, and in fact when the charge is turned off, so does the mass, whose origin is purely electromagnetic. Similar behavior was observed in \cite{Olvera:2019unw} in the study of the Scattering and absorption sections black holes in NED.

\subsection{Dymnikova electric black hole}

In \cite{Dymnikova:2004zc} a SSS solution of NED coupled to gravity that satisfies the weak energy condition and is an electrically charged regular structure, is presented. 

The electrically charged solution was found in the alternative form of NED obtained by the Legendre transformation $P_{\mu \nu}=\mathcal{L}_{F}F_{\mu \nu}$ with $P=P_{\mu \nu}P^{\mu \nu}$ and considering the Hamiltonian-like function $H(P)=2F\mathcal{L}_{F}-\mathcal{L}$. The $P$ frame is related to the $F$ frame by $\mathcal{L}=2PH_{P}-H$ with $H_{P}=dH/dP$.

The proposed function $H(P)$ is;
\begin{equation}
H(P)=\frac{P}{(1+\alpha \sqrt{-P})^2},
\end{equation}

and the function $F$ is given by;

\begin{equation}
F=\frac{P}{(1+\alpha \sqrt{-P})^6},
\end{equation}

where $\alpha =r_{0}^{2}/q \sqrt{2}$, and the electric invariant is $P=-\frac{2q}{r^{4}}$.

The Lagrangian and its derivative are;

\begin{equation}
\mathcal{L}=\frac{2q^{2}(r_{0}^{2}-r^{2})}{(r^{2}+r_{0}^{2})^{3}};\;\;\;\;\;\;\  \mathcal{L}_{F}=\frac{(r^{2}+r_{0}^{2})^{3}}{r^{6}}.
\end{equation}

The line element is given by; 

\begin{equation}
f(r)=g(r)=1-\frac{4M}{r\pi}\left[\arctan\left(\frac{r}{r_0}\right)-\frac{rr_0}{r^2+r_0^2}\right]
\end{equation}

where $r_0=\pi g^2/8M$, $M$ is the mass and $q$ the electric charge. In terms of $q/2M$, a black hole exists for 
$q/2M \leq 0.536$, and for $q/2M > 0.536$ we have an electrically charged self-gravitating particle-like NED structure. The $G_{e}$ and $G_{m}$ factors are given by;

\begin{equation}
G_{m}=1, \quad G_{e}=\left(\frac{\sqrt{2}r^{2}+2q \alpha}{\sqrt{2}r^{2}-4q \alpha}\right)
\end{equation}

where it has been considered that $\mathcal{L}_{FF}=\frac{1}{2}\left(\frac{H_{P}}{F_{P}}-\mathcal{L}_{F}\right)$ (see \cite{Dymnikova:2004zc}).

\begin{figure}[h]
\begin{center}
\includegraphics [width =0.45 \textwidth ]{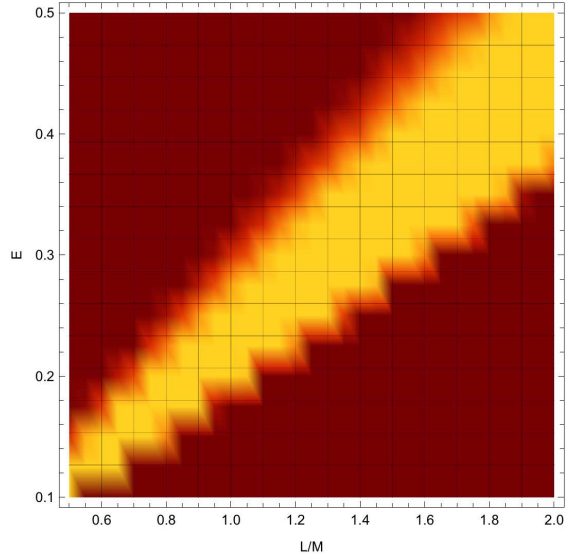}
\end{center}
\caption{Density plot for the parameters $E$ and $L/M$ in Dymnikova BH. Lighter regions represent a higher number of possible orbits.}\label{Fig4}
\end{figure}

Figure (\ref{Fig4}) shows the different values of $E$ and $L/M$ that fulfills the conditions of $E^{2}>0$ , $L^{2}>0$ and $V_{ef}^{''}>0$ for $r=r_{c}$ in the case of Dymnikova magnetic black hole.

\begin{figure}[h]
\begin{center}
\includegraphics [width =0.5 \textwidth ]{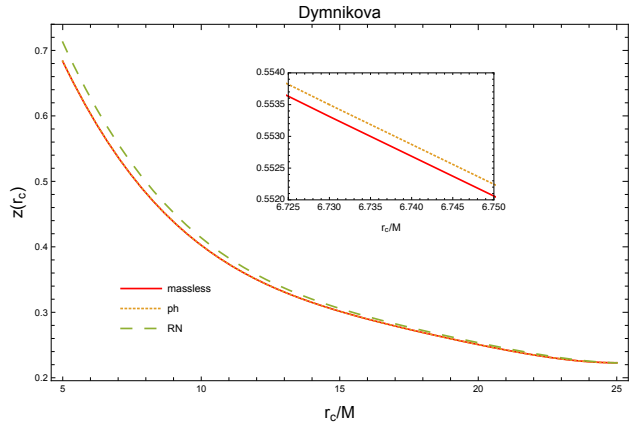}
\end{center}
\caption{Behavior of kinematic redshift for  Dymnikova black hole with values $E=0.5$ and $L/M=2$.}\label{Fig5}
\end{figure}

Choosing the appropriate values observed in the figure (\ref{Fig4}) allows us to obtain the behavior of $z$ for the Dymnikova magnetic black hole. Figure (\ref{Fig5}) shows the $z$ redshift in function of $r_{c}/M$.The behavior of $z$ is similar for photon and massless particles, the $z$ for RN is higher throughout the range considered. In the three cases $z$ decreases as the values of $\frac{r_{c}}{M}$ increase. Analyzing with more detail it is possible to observe that $ z_{massless}< z_{ph}<z_{RN}$ , in this case the effect is more remarkable for RN while for Bardeen BH, massless particles are the dominant.

\begin{figure}[h]
\begin{center}
\includegraphics [width =0.5 \textwidth ]{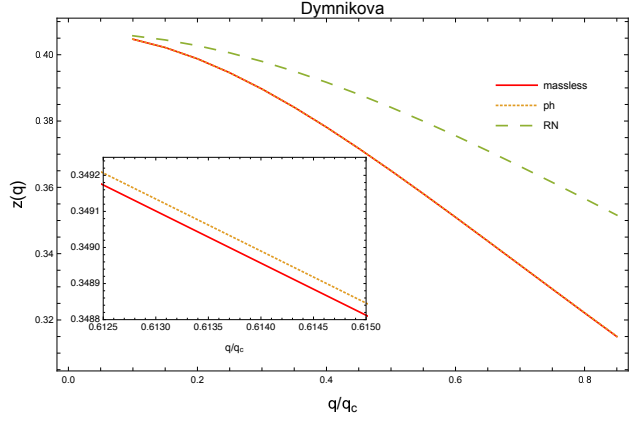}
\end{center}
\caption{Behavior of kinematic redshift for  Dymnikova black hole with values $E=0.5$ and $L/M=2$.}\label{Fig6}
\end{figure}

Now, in figure  (\ref{Fig6}) the behavior of the $z$ redshift as function of $q/q_{c}$ is shown, where $q_{c}=2(0.536)M$. The resemblance of the behavior of photon and massless particles with RN occurs when the charge approaches zero, $z$ decreases as $q/q_{c}$ increases and $ z_{massless}< z_{ph}<z_{RN}$

\subsection{Bronnikov magnetic black hole}

The NLEM Lagrangians coupled to gravity  were  analyzed focusing on the properties that lead to nontrivial regular metrics. Reference \cite{Bronnikov:2000vy} gives one example of a regular magnetic black hole, with the Lagrangian:

\begin{equation}
\mathcal{L}(F)=F {\rm sech}^2 [a(F/2)^{1/4}],
\end{equation}
where $a$ is a constant. The metric function in the line element of the form (\ref{sss}) is

\begin{equation}\label{Bronnokovmetric}
f(r)= g(r)=1-\frac{g^{3/2}}{ar} \left[ {1- \tanh \left( \frac{a \sqrt{g}}{r}\right)} \right],
\end{equation}
where the constant $a$ is related to the mass $m$ and the magnetic charge $g$ by $a=g^{3/2}/(2M)$. The solution corresponds to a BH if $M/g > 0.96$. The electromagnetic invariant  is $F=2g^{2}/r^{4}$. For a purely magnetic charge $G_{e}=1$ and

\begin{equation}
 G_{m}=\frac{g^{2}\sinh^{2}(\frac{g^{2}}{2Mr})\left(-2g^{2}+g^{2}\cosh(\frac{g^{2}}{2Mr})-5Mr\sinh(\frac{g^{2}}{2Mr})\right)}{4Mr\left(-4Mr+g^{2}\tanh(\frac{g^{2}}{2Mr})\right)}
\end{equation}

\begin{figure}[h]
\begin{center}
\includegraphics [width =0.5 \textwidth ]{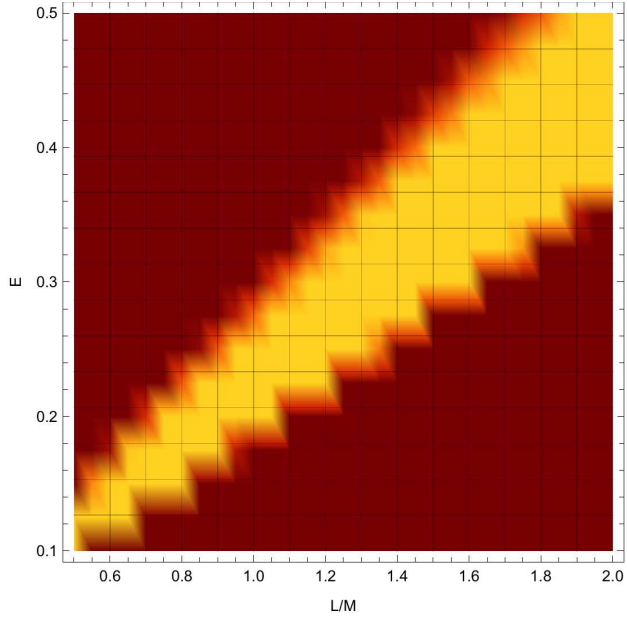}
\end{center}
\caption{Density plot for the parameters $E$ and $L/M$ in Bronnikov BH. Lighter regions represent a higher number of possible orbits.}\label{Fig7}
\end{figure}

As in the previous cases, figure (\ref{Fig7}) shows the ranges of $E$ and $L/M$ for which we can have circular stable orbits. It is possible to observe that in the case of Bardeen black hole, the ranges of values of $E$ and $L/M$ are wider than for Bronnikov and Dymnikova black holes, then the chances of having massive particles that move around of a black hole are higher for Bardeen black hole.

\begin{figure}[h]
\begin{center}
\includegraphics [width =0.5 \textwidth ]{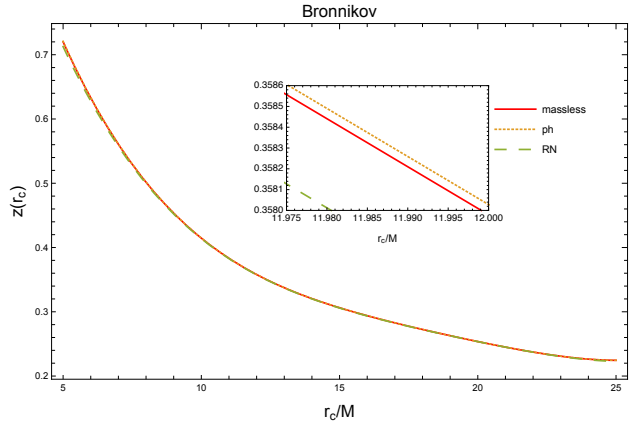}
\end{center}
\caption{Behavior of kinematic redshift for  Bronnikov black hole with values $E=0.5$ and $L/M=2$.}\label{Fig8}
\end{figure}

In figure (\ref{Fig8}) the behavior of the $z$ redshift is showed.  In the same way that Bardeen and Dymnikova black holes the $z$ for Bronnokov black hole decreases as $\frac{r_{c}}{M}$ augments, but the differences while considering photons and massless particles are minimal, in the same way, the same conclusion arises when comparing with RN, however, it is possible to observe this differences by normalizing the different $z$, then it is possible to obtain $ z_{RN}< z_{massless}<z_{ph}$.

\begin{figure}[h]
\begin{center}
\includegraphics [width =0.5 \textwidth ]{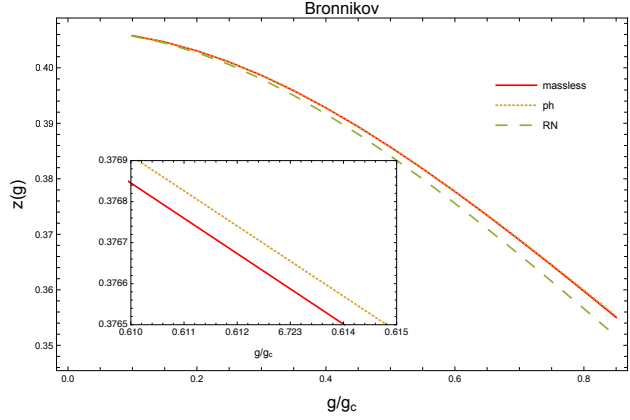}
\end{center}
\caption{Behavior of kinematic redshift for  Bronnikov black hole with values $E=0.5$ and $L/M=2$.}\label{Fig9}
\end{figure}

In figure  (\ref{Fig9}) the behavior of the $z$ redshift as function of $g/g_{c}$ shows that $z$ decreases as $q/q_{c}$ augments. In the case of photon and massless particles, the differences are minimal but the different with RN can be observed. We consider $g_{c}=M/0.96$.

\subsection{Born-Infeld black hole}

The Einstein-Born- Infeld (EBI) generalization of the Reissner Nordstr\"{o}m (RN) BH was obtained by Garc\'ia, Salazar and Pleba\~nski  in \cite{GarciaD.1984}. 

The Einstein-Born-Infeld solution is singular at the origin and it is characterized by three parameters: mass, charge and the BI parameter $b$ that is the maximum attainable electromagnetic field.

The Lagrangian of the  Born$-$Infeld solution is given by;

\begin{equation}\label{BIlagr}
\mathcal{L}(F)=4b^{2}\left(-1+\sqrt{1+\frac{F}{2b^{2}}}\right).
\end{equation}

The line element of the Born$-$ Infeld black hole  is;
\begin{equation}
f(r)=g(r) = 1-\frac{2M}{r}+\frac{2}{3}r^{2}b^{2}\left(1-\sqrt{1+\frac{q^{2}}{b^{2}r^{4}}}\right)+\frac{2}{3}\frac{q^{2}}{r}\sqrt{\frac{b}{q}}
\mathbb{F}\left(\arccos\left[\frac{br^{2}/q-1}{br^{2}/q+1}\right],\frac{1}{\sqrt{2}}\right),
\end{equation}

where $\mathbb{F}$ is the elliptic integral of the first kind, $M$ is the mass parameter, $q$ is the electric charge (both in length units) and $b$ is the Born-Infeld parameter that corresponds to the magnitude of the electric field at $r=0$. The nonvanishing component of the electromagnetic field is $F_{rt}=q\left(r^{4}+\frac{q^{2}}{b^{2}}\right)^{-1/2}$. The nonlinear factors $G_m$ and $G_e$, in case that  only electric charge is considered, $q_m=0$ and $q_{e}=q$ ;

 \begin{equation}
G_{m}=1, \quad G_{e}=\left(1+\frac{q^{2}}{b^{2}r_c^{4}}\right).
\end{equation}

\begin{figure}[h]
\begin{center}
\includegraphics [width =0.5 \textwidth ]{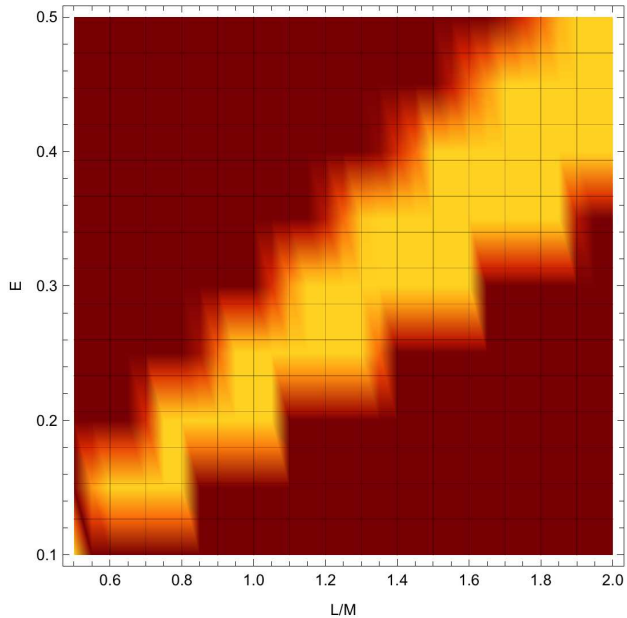}
\end{center}
\caption{Density plot for the parameters $E$ and $L/M$ in Born$-$Infeld BH with $b=0.3$. Lighter regions represent a higher number of possible orbits.}\label{Fig10}
\end{figure}

The values of $E$ and $L/M$ that can be considered to obtain circular stable orbits are shown in figure (\ref{Fig10}). 
The ranges of values for obtaining stable orbits are not very wide, the geodesic structure of the EBI BH was studied in \cite{Breton_2002} where the ranges for different orbits are discussed.

As in the case of Bronnikov black hole the $z$ redshift for massless particles and photon do not present a noticeable difference for Born-Infeld black hole. However, in the two cases, $z$ decreases as $\frac{r_{c}}{M}$ augments. In the same way, the same conclusion arises when comparing with RN, however, it is possible to compare these differences by normalizing the $z$ of Born-Infeld black hole with respect to RN, obtaining that $ z_{massless}< z_{RN}<z_{ph}$. This relation for $z$ is independent of the parameter $b$, however, by increasing it, behavior for photons and massless particles gets closer to RN.

 \begin{figure}[h]
\begin{center}
\includegraphics [width =0.5 \textwidth ]{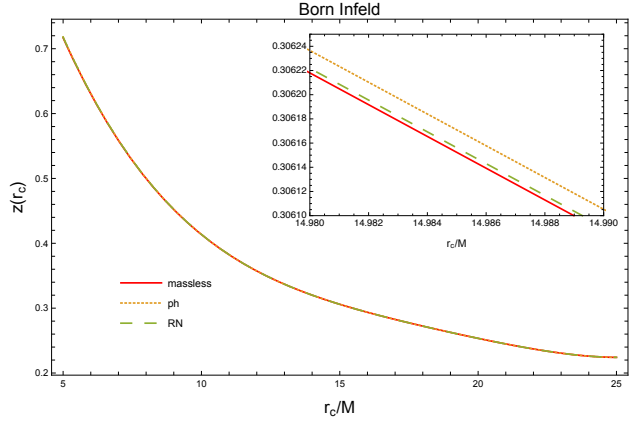}
\end{center}
\caption{Behavior of kinematic redshift for  Born$-$Infeld black hole  with values $E=0.5$, $b=0.3$ and $L/M=2$.}\label{Fig11}
\end{figure}

\begin{figure}[h]
\begin{center}
\includegraphics [width =0.5 \textwidth ]{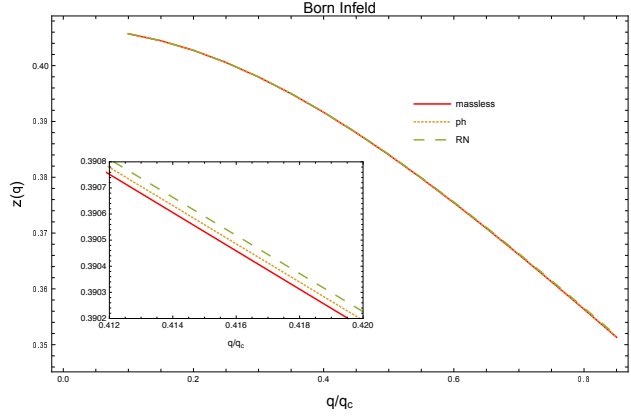}
\end{center}
\caption{Behavior of kinematic redshift for  Born$-$Infeld black hole  with values $E=0.5$, $b=0.3$ and $L=2$.}\label{Fig12}
\end{figure}

In figure (\ref{Fig12}) it can be observed that $ z_{massless}< z_{ph}<z_{RN}$, when we consider $z$ redshift in function of $q/q_{c}$ where $q_{c}=M$. 

\subsection{Asymptotic behaviour}

The asymptotic behavior of $z$ would depend on the behavior of the metric functions and the magnetic and electric factors. When considering the effects of NED for the different black holes the behavior of $z$ when $r_{c}/M \gg 1$ is the following:

\begin{itemize}
\item Reissner-Nordstr\"{o}m  black hole

\begin{equation}
z \approx \sqrt{\frac{M}{r_{c}}}+\left(\frac{5}{2}-\frac{0.5q^{2}}{M^{2}}\right)\left(\frac{M}{r_{c}}\right)^{3/2}+\cdots
\end{equation}

\item Bardeen black hole

\begin{equation}
z \approx \sqrt{\frac{2}{3}}\sqrt{\frac{M}{r_{c}}}+\frac{5}{\sqrt{6}}\left(\frac{M}{r_{c}}\right)^{3/2}+\cdots
\end{equation}

\item Dymnikova black hole

\begin{equation}
z \approx \sqrt{\frac{M}{r_{c}}}+\left(\frac{5}{2}-\frac{0.88q^{2}}{M^{2}}\right)\left(\frac{M}{r_{c}}\right)^{3/2}+\cdots
\end{equation}

\item Bronnikov black hole

\begin{equation}
z \approx \sqrt{\frac{M}{r_{c}}}+\left(\frac{5}{2}-\frac{0.424g^{2}}{M^{2}}\right)\left(\frac{M}{r_{c}}\right)^{3/2}+\cdots
\end{equation}

\item Born-Infeld black hole

\begin{equation}
z \approx \sqrt{\frac{M}{r_{c}}}+\left(\frac{5}{2}-\frac{0.5q^{2}}{M^{2}}\right)\left(\frac{M}{r_{c}}\right)^{3/2}+\cdots
\end{equation}
\end{itemize}

It is possible to observe that the behavior of $z$ is the same when considering first-order for Dymnikova, Born-Infeld, Bronnikov, and RN black holes. In the case of Bardeen black hole, the difference is given by the factor $\sqrt{2/3}$ that describes best the NED's contribution in the other cases, said factor tends to one in the proposed limit. At second-order, small differences are already observed but the behavior of RN and BI is the same and does not depend on the Born-Infeld parameter.

\section{Conclusions}

This paper  shows how the red-blueshifts of the photons emitted by massless test particles  that move around a black hole are modified by considering the effective metric generated by the action of NLED. The kinematic shifts for photons are modified by the magnetic and electric factors that make the difference between the linear and non-linear electrodynamics. 
While the NLED effects are not manifest on the behavior of red-blueshifts of the photons, they are present when considering the differences between massless test particles and photons, also when its redshifts are compared with the redshifts of Reissner-Nordstr\"{o}m black hole. The redshifts from two magnetic black holes and two electric black holes were studied and analyzed. 

In the cases of the Bardeen and Bronnikov black holes that have magnetic charge, the behavior of $z$ for both black holes decreases as $\frac{r_{c}}{M}$ augments. For the Bardeen black hole, the difference between massless test
particles an photons is appreciable but the difference between RN and massless test particles is indistinguishable prima facie. For Bronnikov black hole the differences while considering photons, massless test particles and RN are indistinguishable, however, it is possible to study these differences by normalizing the different $z$.
The behavior of $z$ for both black holes decreases as $g/g_{c}$ augments, the difference between massless test particles, photons, and RN is appreciable in the case of the Bardeen black hole, for Bronnikov black hole the differences are not easily distinguishable.

For the electric black holes of Born-Infeld and Dymnikova, the redshifts have the same behavior that in the case of the magnetic black holes, i.e $z$ decreases as $\frac{r_{c}}{M}$  and $q/q_{c}$ increase. For  the Born-Infeld black hole the differences of $z$ between photons, RN and massless test particles are closer. 

It is worth mentioning that for higher values of energy and angular momentum, more stable orbits appear around the black hole. However, as can be observed from the density plots (figures \ref{Fig1}, \ref{Fig4}, \ref{Fig7} and \ref{Fig10}) the condition $(L/M)/E>1$ must  be fulfilled in order to obtain the required orbits. The ratio $(L/M)/E$ from which a considerable number of circular orbits appear (lighter region in figures \ref{Fig1}, \ref{Fig4}, \ref{Fig7} and \ref{Fig10}) varies depending on the BH and its parameters.

\bibliographystyle{unsrt}

\bibliography{bibliografia}

\begin{thebibliography}{10}

\bibitem{Begelman1898}
Mitchell~C. Begelman.
\newblock Evidence for black holes.
\newblock {\em Science}, 300(5627):1898--1903, 2003.

\bibitem{PMID:16267548}
Zhi-Qiang Shen, KY~Lo, M-C Liang, Paul T~P Ho, and J-H Zhao.
\newblock A size of $\sim$1 au for the radio source sgr a* at the centre of the
  milky way.
\newblock {\em Nature}, 438(7064):62—64, November 2005.

\bibitem{2009ASSP....9..361E}
F.~{Eisenhauer}, G.~{Perrin}, W.~{Brandner}, C.~{Straubmeier}, A.~{B{\"o}hm},
  H.~{Baumeister}, F.~{Cassaing}, Y.~{Cl{\'e}net}, K.~{Dodds-Eden},
  A.~{Eckart}, E.~{Gendron}, R.~{Genzel}, S.~{Gillessen}, A.~{Gr{\"a}ter},
  C.~{Gueriau}, N.~{Hamaus}, X.~{Haubois}, M.~{Haug}, T.~{Henning},
  S.~{Hippler}, R.~{Hofmann}, F.~{Hormuth}, K.~{Houairi}, S.~{Kellner},
  P.~{Kervella}, R.~{Klein}, J.~{Kolmeder}, W.~{Laun}, P.~{L{\'e}na},
  R.~{Lenzen}, M.~{Marteaud}, V.~{Naranjo}, U.~{Neumann}, T.~{Paumard},
  S.~{Rabien}, J.~R. {Ramos}, J.~M. {Reess}, R.~R. {Rohloff}, D.~{Rouan},
  G.~{Rousset}, B.~{Ruyet}, A.~{Sevin}, M.~{Thiel}, J.~{Ziegleder}, and
  D.~{Ziegler}.
\newblock {GRAVITY: Microarcsecond Astrometry and Deep Interferometric Imaging
  with the VLT}.
\newblock {\em Astrophysics and Space Science Proceedings}, 9:361, January
  2009.

\bibitem{Herrera-Aguilar:2015kea}
Alfredo Herrera-Aguilar and Ulises Nucamendi.
\newblock {Kerr black hole parameters in terms of the redshift/blueshift of
  photons emitted by geodesic particles}.
\newblock {\em Phys.\ Rev.\ D}, 92(4):045024, 2015.

\bibitem{Becerril:2016qxf}
Ricardo Becerril, Susana Valdez-Alvarado, and Ulises Nucamendi.
\newblock {Obtaining mass parameters of compact objects from redshifts and
  blueshifts emitted by geodesic particles around them}.
\newblock {\em Phys.\ Rev.\ D}, 94(12):124024, 2016.

\bibitem{Kraniotis:2019ked}
G.V. Kraniotis.
\newblock {Gravitational redshift/blueshift of light emitted by geodesic test
  particles, frame-dragging and pericentre-shift effects, in the Kerr-Newman-de
  Sitter and Kerr-Newman black hole geometries}.
\newblock 12 2019.

\bibitem{Gutierrez:1981ed}
S.~A. Gutierrez, A.~L. Dudley, and J.~F. Plebanski.
\newblock {Signals and Discontinuities in General Relativistic Nonlinear
  Electrodynamics}.
\newblock {\em J. Math. Phys.}, 22:2835--2848, 1981.

\bibitem{Breton:2016mqh}
N.~Breton and L.A. Lopez.
\newblock {Quasinormal modes of nonlinear electromagnetic black holes from
  unstable null geodesics}.
\newblock {\em Phys.\ Rev.\ D}, 94(10):104008, 2016.

\bibitem{Olvera:2019unw}
J.C. Olvera and L.A. López.
\newblock {Scattering and absorption sections of nonlinear electromagnetic
  black holes}.
\newblock {\em Eur. Phys. J. Plus}, 135(3):288, 2020.

\bibitem{Stuchlik:2019uvf}
Zdenek Stuchlík and Jan Schee.
\newblock {Shadow of the regular Bardeen black holes and comparison of the
  motion of photons and neutrinos}.
\newblock {\em Eur. Phys. J. C}, 79(1):44, 2019.

\bibitem{Allahyari:2019jqz}
Alireza Allahyari, Mohsen Khodadi, Sunny Vagnozzi, and David~F. Mota.
\newblock {Magnetically charged black holes from non-linear electrodynamics and
  the Event Horizon Telescope}.
\newblock {\em JCAP}, 02:003, 2020.

\bibitem{Cuzinatto:2015xta}
R.R. Cuzinatto, C.A.M. Melo, K.C. Vasconcelos, L.G. Medeiros, and P.J. Pompeia.
\newblock {Non-linear effects on radiation propagation around a charged compact
  object}.
\newblock {\em Astrophys. Space Sci.}, 359(2):59, 2015.

\bibitem{AyonBeato:2000zs}
Eloy Ayon-Beato and Alberto Garcia.
\newblock {The Bardeen model as a nonlinear magnetic monopole}.
\newblock {\em Phys. Lett.}, B493:149--152, 2000.

\bibitem{Dymnikova:2004zc}
Irina Dymnikova.
\newblock {Regular electrically charged structures in nonlinear electrodynamics
  coupled to general relativity}.
\newblock {\em Class. Quant. Grav.}, 21:4417--4429, 2004.

\bibitem{Bronnikov:2000vy}
Kirill~A. Bronnikov.
\newblock {Regular magnetic black holes and monopoles from nonlinear
  electrodynamics}.
\newblock {\em Phys. Rev.}, D63:044005, 2001.

\bibitem{GarciaD.1984}
A.~Garc{\'i}a~D., H.~Salazar~I., and J.~F. Pleba{\'{n}}ski.
\newblock Type-d solutions of the einstein and born-infeld
  nonlinear-electrodynamics equations.
\newblock {\em Il Nuovo Cimento B (1971-1996)}, 84(1):65--90, 1984.

\bibitem{Breton_2002}
Nora Bret{\'{o}}n.
\newblock Geodesic structure of the born{\textendash}infeld black hole.
\newblock {\em Classical and Quantum Gravity}, 19(4):601--612, jan 2002.

\end{thebibliography}

\end{document}